\documentclass[
  journal=largetwo,
  manuscript=article-type,
  year=2020,
  volume=37,
]{cup-journal}

\usepackage{amsmath}
\usepackage[nopatch]{microtype}
\usepackage{booktabs}

\title{NuSTAR Observations of the Galaxy Cluster Abell 3667}

\author{Mohammad S. Mirakhor}
\affiliation{Department of Physics and Astronomy, The University of Alabama in Huntsville, 301 Sparkman Drive, Huntsville, AL 35899, USA}
\email[Mohammad S. Mirakhor]{msm0033@uah.edu}

\author{Stephen A. Walker}
\affiliation{Department of Physics and Astronomy, The University of Alabama in Huntsville, 301 Sparkman Drive, Huntsville, AL 35899, USA}


\begin{document}

\begin{abstract}
We present an analysis of the hard X-ray emission from the central region of Abell 3667 using deep \textit{NuSTAR} observations. While previous studies on the nature of the hard X-ray excess have been controversial, our analysis of the central region suggests that the excess is primarily thermal, best described by a two-temperature (2T) model, with the high-temperature component likely arising from merger-induced heating. This interpretation contrasts with some earlier suggestions of non-thermal emission due to inverse Compton scattering of relativistic electrons. Additionally, we set a lower limit on the magnetic field strength of $\sim 0.2 \, \mu$G in the central region, consistent with values found in other dynamically active clusters and compatible with those inferred from equipartition and Faraday rotation measurements. Since our study is focused on the central region of the cluster, further high-resolution observations of the outer regions will be critical to fully disentangle the thermal and non-thermal contributions to the X-ray. 
\end{abstract}


\section{Introduction}
\label{sec:intro}

The intracluster medium (ICM) of galaxy clusters is known to harbour a significant population of relativistic particles and large-scale magnetic fields, predominantly revealed through radio observations \citep[see][for a recent review]{vanWeeren2019}. These magnetic fields, typically on the order of a few $\mu$Gauss, facilitate the emission of synchrotron radiation from radio-emitting relativistic electrons, leading to the formation of large-scale, diffuse structures known as radio haloes and relics. Furthermore, these relativistic electrons can enhance the energies of cosmic microwave background (CMB) photons via inverse Compton (IC) scattering, creating a non-thermal high-energy tail in the X-ray spectrum of galaxy clusters \citep[e.g.][]{Rephaeli1979,Sarazin2000}.

\begin{figure*}
\begin{center}
\includegraphics[width=0.45\textwidth]{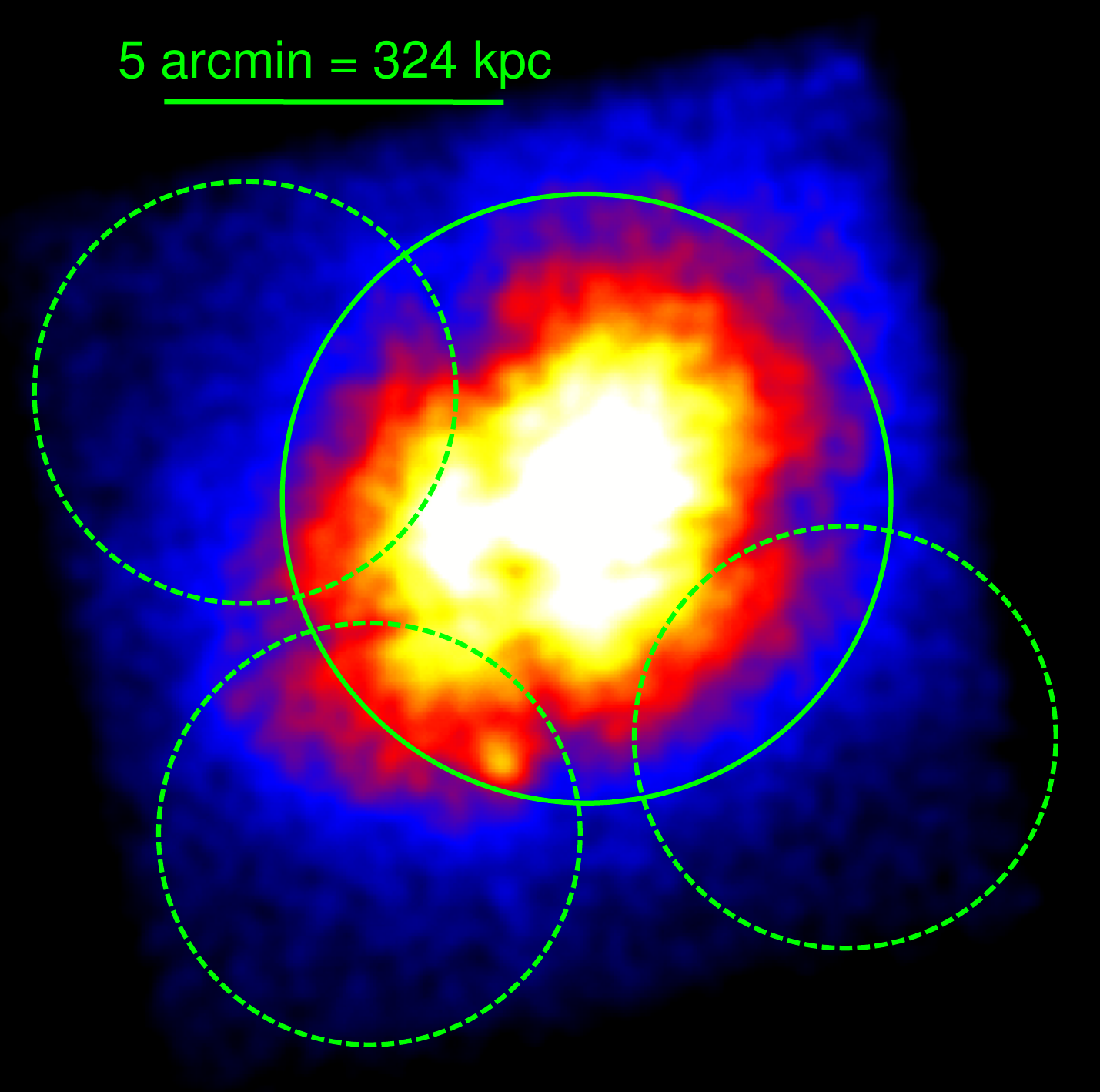} 
\includegraphics[width=0.45\textwidth]{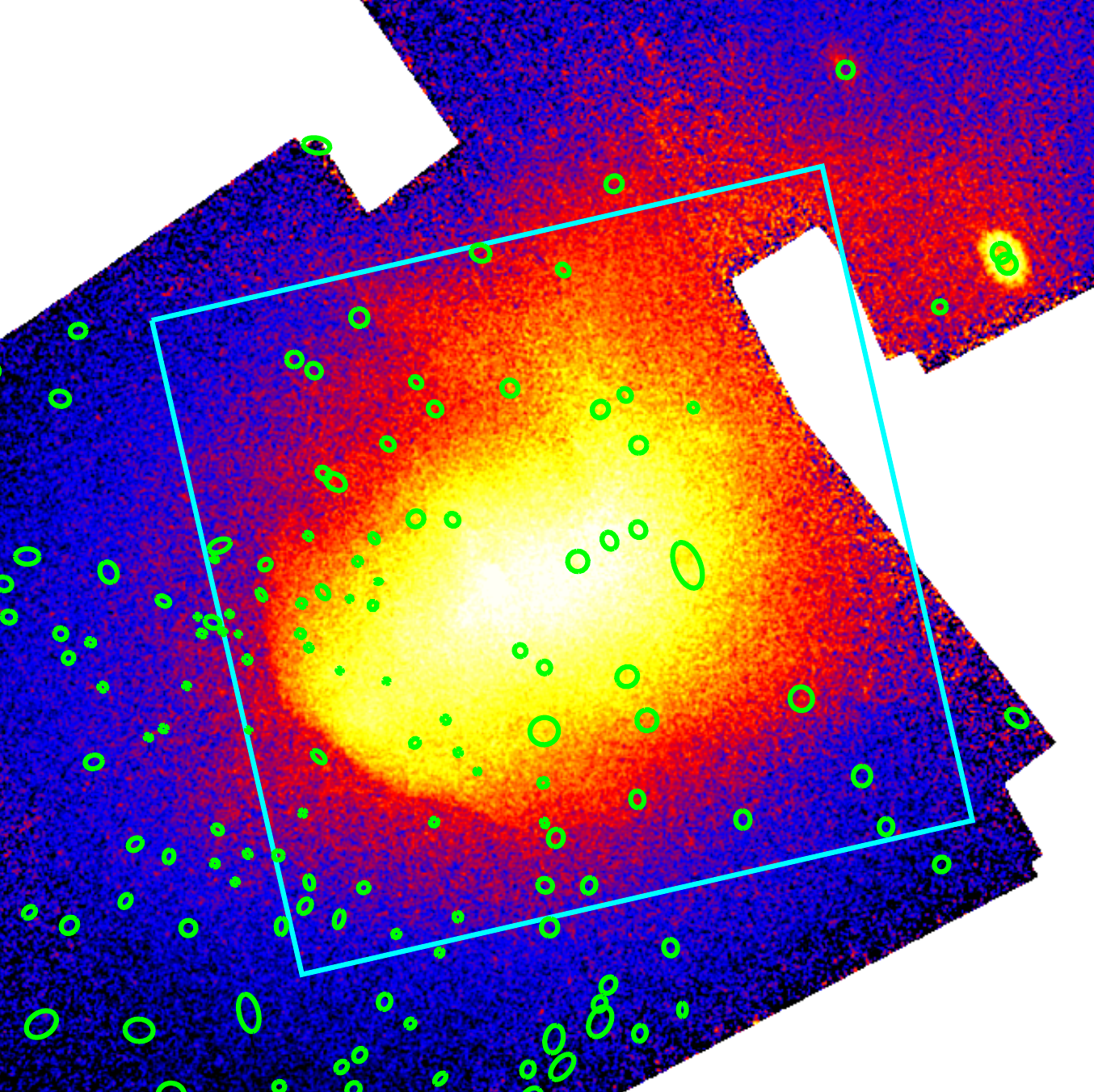} 
\end{center}
\caption{\textit{Left}: Background-subtracted and exposure-corrected \textit{NuSTAR} mosaic image of the galaxy cluster Abell 3667 in the 4–24 keV energy band, combining data from both telescopes. The image was smoothed using a Gaussian kernel with a $\sigma = 17.2$ arcsec (7 pixels) to match \textit{NuSTAR}'s PSF of $\sim 18$ arcsec FWHM. The solid green circle (4.5 arcmin radius) indicates the region from which the source spectra were extracted, while the dashed green circles mark the regions used to extract the background spectra. The overlapping areas are excluded from the background regions. \textit{Right}: Mosaicked Chandra image of Abell 3667, showing the locations of the point sources. The cyan box outlines the extent of our \textit{NuSTAR} observations.}
\label{fig: A3667_img}
\end{figure*}

The detection of non-thermal radiation is crucial for directly estimating the mean magnetic field strength in the ICM, circumventing the need for assumptions about energy equipartition between the magnetic fields and relativistic electrons \citep{Rephaeli2008}. Magnetic fields significantly influence particle acceleration processes and affect the thermal gas's energy distribution by modifying heat conduction, large-scale gas motions, and cosmic ray propagation \citep[e.g.][]{Carilli2002,vanWeeren2019}. Cosmological simulations suggest that magnetic fields can have a profound impact on the dynamics and structure of the ICM, particularly in sloshing cool-core clusters, where local magnetic field amplification can reach near-equipartition with the thermal pressure of the ICM \citep[e.g.][]{ZuHone2011}. Consequently, measuring non-thermal emissions can provide insights into whether the average magnetic field strength in the ICM is sufficient to influence the dynamical state of the thermal gas.

However, the observational detection of non-thermal IC emission presents significant challenges, especially below 10 keV, where thermal X-ray photons are predominant \citep{Rephaeli2008}. At higher energies, where thermal Bremsstrahlung emission declines steeply, non-thermal IC emission may be detectable as a hard X-ray excess superimposed on the thermal spectrum of clusters.

The quest for non-thermal emission began with \textit{HEAO-1} data analyses \citep{Rephaeli1987,Rephaeli1988}, but the first claim of non-thermal emission was obtained from deep observations of the Coma cluster using \textit{RXTE} and \textit{Beppo-SAX} \citep{Rephaeli1999,Fusco-Femiano1999}. Although several claims of non-thermal IC emission above 20 keV have been made in other galaxy clusters, these detections have often been marginally significant and contentious \citep[see][for a review]{Rephaeli2008}. Follow-up observations with the \textit{Suzaku} and \textit{Swift} satellites largely failed to support the earlier claims of non-thermal IC emission detected by RXTE and Beppo-SAX \citep[e.g.][]{Ajello2009,Wik2012,Ota2014}, with the Bullet cluster and Abell 3667 being a notable exception \citep{Ajello2010}. However, recent findings from \textit{NuSTAR}, the first focusing high-energy X-ray observatory operating between 3 and 79 keV, suggest that the global spectrum of the Bullet cluster can be explained by a multi-temperature thermal model, indicating that its hard X-ray emission likely originates from thermal processes \citep{Wik2014}.

\begin{figure*}[ht]
\begin{center}
\includegraphics[width=\textwidth]{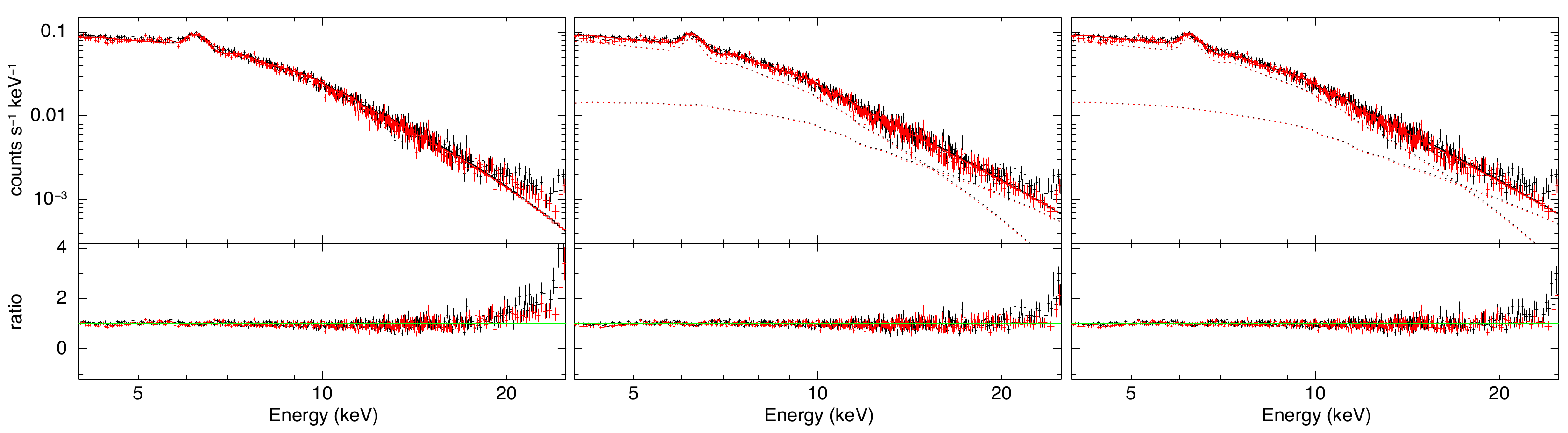} 
\end{center}
\caption{\textit{Upper panels}: Global spectra of both telescopes of Abell 3667, extracted from a circular region of radius 4.5 arcmin shown in the left-hand panel of Fig. \ref{fig: A3667_img}. The spectra were fitted to a single-temperature model (\textit{left}), two-temperature model (\textit{middle}), and single-temperature plus power-law model with a free photon index (\textit{right}). The solid lines are the best-fitting models to the spectra. The components of the two-temperature and single-temperature plus power-law models are shown as dashed lines, with the IC component (\textit{right}) dominating at energies above $\sim 10$ keV. \textit{Lower panels}: The ratio of the data to the best-fitting models. The two-temperature model provides a better fit to the spectral data than the model that includes the power-law component.}
\label{fig: A3667_fitting}
\end{figure*}

Currently, the exploration for non-thermal IC emission using \textit{NuSTAR} is confined to a limited number of galaxy clusters, including the Bullet cluster \citep{Wik2014}, the Coma cluster \citep{Gastaldello2015}, Abell 523 \citep{Cova2019}, Abell 2163 (\citep{Rojas2021}, CL 0217+70 \citep{Tumer2022CL0217}, and SPT-CL J2031--4037 \citep[][]{mirakhor2022possible}. In this paper, we leverage \textit{NuSTAR}’s remarkable capability to focus X-rays in the hard energy band above 10 keV to investigate the nature of the hard emission in the merging galaxy cluster Abell 3667, using our new deep observations totalling 208 ks. Abell 3667 is a prominent galaxy cluster situated at a redshift of approximately 0.055. It ranks among the brightest X-ray sources in the southern sky, exhibiting a luminosity of $5.1 \times 10^{44}$ erg s$^{-1}$ in the 0.4--2.4 keV band \citep{edge1990x}. Its intricate morphology and notable hard excess emission have made it a focal point for studies across various wavelengths. The investigation of hard X-ray emission in Abell 3667 has garnered considerable attention, particularly regarding the underlying mechanisms driving the excess emission.

\textit{NuSTAR} employs two coaligned X-ray telescopes, designated FPMA and FPMB, which feature an effective area of $2 \times 350$ cm$^2$ at 9 keV and imaging half-power diameter of 58 arcsec. While its effective area may be lower than that of previous instruments, its focusing capability significantly reduces point source contamination and background noise. \textit{NuSTAR} is particularly valuable for studying galaxy clusters and their associated hard excess emission. Its ability to focus X-rays in the hard energy range allows for detailed investigations of the high-energy processes occurring in these clusters. By providing high-resolution images and spectral data, \textit{NuSTAR} enables us to probe the physical mechanisms behind this excess emission, enhancing our understanding of the dynamics and evolution of galaxy clusters in the universe.

\section{Observations and data processing}
\label{sec: data}

The galaxy cluster Abell 3667 was observed by \textit{NuSTAR} with an unfiltered exposure totalling 208 ks (PI: M. S. Mirakhor) on June 23, 2024. Datauction was performed using standard pipeline processing tools (\textsc{HEASoft} v6.29 and \textsc{Nustardas} v2.1.1) along with the 20240715 version of the \textit{NuSTAR} Calibration Database. Data processing began with the execution of the \texttt{nupipeline} script to create cleaned event files. We screened the data for periods affected by Earth occultations and increased background caused by the South Atlantic Anomaly (SAA). To enforce stringent criteria regarding SAA occurrences, we set the \texttt{saamode} parameter to STRICT in the \texttt{nupipeline} script. Additionally, the '\texttt{tentacle}' parameter was activated to flag time intervals when the CZT detectors recorded heightened event rates upon entering the SAA region. The filtered exposure times for the FPMA and FPMB telescopes are 157.9 ks and 161.6 ks, respectively.

From the cleaned event files, we generated images, light curves, and spectra using the \texttt{nuproducts} script. Exposure maps were created with \texttt{nuexpomap}, accounting for vignetting effects. While executing \texttt{nuproducts}, we set the '\texttt{extended}' parameter to '\texttt{yes}' to produce response matrix and auxiliary response files suitable for extended sources. Light curves were constructed in the 1.6–20 keV (solar) and 50–160 keV (particle) bands with 100 s bins, and periods of excess count rates were manually identified and excluded from the Good Time Interval (GTI) files.

The background for \textit{NuSTAR}, which varies spectrally and spatially within the field of view, can generally be divided into four primary components \citep{Wik2014}: the first is an instrumental background that predominates at energies above approximately 20 keV; the second is an "aperture" background from stray light due to unfocused cosmic X-rays leaking through the aperture stops; the third involves reflected and scattered stray light, mainly from cosmic X-rays, the Earth, and the Sun, impacting the detectors because of the observatory's open geometry; and the fourth is a focused cosmic background from unresolved sources within the field of view, contributing significantly below 15 keV. In this analysis, we used the \textsc{idl} routines \texttt{nuskybgd} recommended in \citet{Wik2014} to characterise the background and create scaled background spectra for the desired regions and energy bands. Above 20 keV, the background primarily arises from instrumental fluorescence and cosmic ray activation during SAA passages, alongside the instrumental particle background continuum. Between 5–20 keV, the aperture component is the dominant background source, while the solar component, combined with the model of the instrumental lines, significantly impacts the spectrum below 5 keV. The focused cosmic background contributes primarily below 15 keV, lying beneath the aperture lines. For a comprehensive explanation of \textit{NuSTAR} background modeling, we refer to \citet{Wik2014}.

Figure \ref{fig: A3667_img} (left) presents the background-subtracted and exposure-corrected mosaicked image of the Abell 3667 cluster in the 4–24 keV energy band. The image has been smoothed using a Gaussian kernel with a $\sigma$ of 17.2 arcseconds (7 pixels) to align with \textit{NuSTAR}'s point spread function (PSF), which has a full width at half-maximum (FWHM) of approximately 18 arcseconds.

\section{Spectral analysis}
\label{sec: spectra}

\begin{table*}
\begin{center}  
\caption {Results of the spectral analysis of Abell 3667. The statistical uncertainties are at the 90 per cent confidence levels, followed by the 90 per cent systematic uncertainties.}
\resizebox{\textwidth}{!}{ 
\begin{tabular}{lcccccccc}
\hline
    Model  & $T_1$  & $Z$ & Norm$_1$ & $T_2$ & $\Gamma$ & Norm$_2$ & IC flux & C-stat/dof   \\
          & (keV) &  (Z$_\odot$)  & ($10^{-2}$ cm$^{-5}$)  & (keV)   & & ($10^{-3}$ cm$^{-5}$) & ($10^{-12}$ erg s$^{-1}$ cm$^{-2}$) &   \\    \hline
1T & $9.1_{-0.1-0.1}^{+0.1+0.1}$ & $0.25_{-0.02-0.01}^{+0.02+0.01}$ & $2.28_{-0.03-0.03}^{+0.03+0.04}$\footnote{\parbox{\linewidth}{\label{note1}Normalization of the first APEC thermal component for the FPMA instrument.}} &  &   &  & & $4465.3_{-24}^{+32}/3694$ \\
   &  &  & $2.23_{-0.03-0.04}^{+0.03+0.04}$\footnote{\parbox{\linewidth}{\label{note2}Normalization of the first APEC thermal component for the FPMB instrument.}} &  &   &  & & \\
 
2T & $5.8_{-0.3-0.3}^{+0.3+0.4}$ & $0.24_{-0.02-0.02}^{+0.02+0.03}$  & $2.10_{-0.07-0.11}^{+0.10+0.12}$\footref{note1}  & $25.8_{-2.1-3.1}^{+6.1+7.8}$ &  & $6.23_{-0.79-0.77}^{+0.50+0.52}$\footnote{\parbox{\linewidth}{\label{note3}Normalization of the second APEC thermal component for the FPMA instrument.}} &  & $4211.2_{-20}^{+23}/3691$ \\ 
 &  &   & $2.09_{-0.07-0.11}^{+0.10+0.11}$\footref{note2}  &  &  & $6.21_{-0.89-0.44}^{+0.35+0.40}$\footnote{\parbox{\linewidth}{\label{note4}Normalization of the second APEC thermal component for the FPMB instrument.}} &  &  \\ 

1T$+$IC  & $6.6_{-0.2-0.3}^{+0.2+0.4}$ & $0.25_{-0.02-0.04}^{+0.03+0.03}$ & $2.11_{-0.14-0.12}^{+0.13+0.14}$\footref{note1} &  & $1.6_{-0.2-0.1}^{+0.1+0.1}$ & $1.18_{-0.38-0.25}^{+0.67+0.33}$\footnote{\parbox{\linewidth}{\label{note5}Normalization of the IC component for the FPMA instrument, given in units of photons keV$^{-1}$ cm$^{-2}$ s$^{-1}$ at 1 keV (10$^{-3}$).}} & $9.52_{-5.51-0.32}^{+0.46+0.38}$ & $4230.1_{-20}^{+22}/3691$  \\
  &  &  & $2.13_{-0.13-0.08}^{+0.13+0.08}$\footref{note2} &  & & $1.14_{-0.42-0.20}^{+0.55+0.20}$\footnote{\parbox{\linewidth}{\label{note6}Normalization of the IC component for the FPMB instrument, given in units of photons keV$^{-1}$ cm$^{-2}$ s$^{-1}$ at 1 keV (10$^{-3}$).}} &  &  \\
\hline
\end{tabular}
\label{table: model_comparison}
}
\end{center} 
\end{table*}

The spectrum of the Abell 3667 cluster was extracted from a circular region with a radius of 4.5 arcminutes, centered on the cluster's peak surface brightness (see Fig. \ref{fig: A3667_img}). We excluded data below 4 keV from our analysis based on findings by \citet{Madsen2020}, which indicate that the effective area of the FPMA detector may be unreliable in this energy range. This limitation is due to a tear in the thermal blanket of the \textit{NuSTAR} optics, allowing additional low-energy photons to be detected.

Above 20 keV, the spectra are dominated by background contributions. To account for systematic uncertainties arising from background fluctuations, we followed the procedure outlined in \citet{Wik2014}. Specifically, we simulated 1000 realizations of the background for each spectrum, incorporating systematic errors for various background components: 3 per cent for the instrumental component, 8 per cent for the aperture component, 42 per cent for the focused cosmic component, and 10 per cent for the Solar component. Assuming Gaussian fluctuations, each background realization was subtracted from the source spectrum, and the systematic uncertainties for the spectral models were then estimated.

We initiated the fitting process by simultaneously analysing the spectral data from both telescopes using a single-temperature (1T) model. This model incorporates an APEC thermal component \citep{Smith2001}, with Galactic absorption fixed at \(N_{\rm{H}} = 3.67 \times 10^{20} \, \text{cm}^{-2}\). We initially allowed the redshift to vary freely, then fixed it to the value determined from the fitting to account for offsets between the observed and known redshifts from \textit{NuSTAR}, which are likely due to imperfect gain calibration \citep[see Appendix B for][]{Rojas2023}. The normalisation constants for both instruments were allowed to vary freely to account for any flux or instrumental differences between FPMA and FPMB. The spectral fitting was performed using the widely adopted X-ray spectral fitting package, \textsc{xspec} \citep{Arnaud1996}. To obtain the best-fitting parameters, we employed the C-statistic (C-stat), which utilises the W-statistic, a modification of the Cash statistic. The fitting results yielded a temperature of \(T = 9.1 \pm 0.1 \, \text{keV}\) and a metal abundance of \(Z = 0.25 \pm 0.02 \, Z_\odot\) at 90\% confidence levels. The C-stat value for the 1T model fit is 4465.3, with 3694 degrees of freedom. A summary of the best-fitting parameters is presented in Table \ref{table: model_comparison}, and the fit to the spectra from both telescopes is illustrated in the left panel of Fig. \ref{fig: A3667_fitting}, where a notable excess is observed above 20 keV.

The 1T model we initially considered provides a very basic representation of the spectral data, but it is unlikely to accurately reflect the physical conditions, particularly for clusters with complex structures like Abell 3667. To improve upon this, we adopted a two-temperature (2T) model to better capture the data. This model consists of two thermal components from the APEC plasma code, with their metal abundances tied together. The best-fit values from the 2T model yield temperatures of $T_1 = 5.8_{-0.3}^{+0.3}$ keV and $T_2 = 25.8_{-2.1}^{+6.1}$ keV, along with an abundance of $Z = 0.24 \pm 0.02$ Z$_\odot$, all at 90\% confidence (see Table \ref{table: model_comparison} and the middle panel of Fig \ref{fig: A3667_fitting}). The C-stat value for this fit is 4211.2, with 3691 degrees of freedom, indicating a superior fit relative to the 1T model.

We also performed a spectral fit of the X-ray data using a single-temperature plus power-law model (1T + IC), where the photon spectral index of the power-law component was allowed to vary freely. The best-fitting parameters derived from this model yielded a temperature of \( T = 6.6 \pm 0.2 \, \text{keV} \) and an abundance of \( Z = 0.25_{-0.02}^{+0.03} \, Z_\odot \), with uncertainties quoted at the 90\% confidence level (see Table \ref{table: model_comparison}). The best fit yielded an index of $1.6_{-0.2}^{+0.1}$ and a 20--80 keV flux of \( 9.52_{-5.51}^{+0.46} \times 10^{-12} \, \text{erg} \, \text{s}^{-1} \, \text{cm}^{-2} \). Our results indicate a noticeable improvement in the fit statistics when a power-law component is added to the single-temperature model. However, when comparing the 1T + IC model to the 2T model, our analysis reveals that the 2T model provides a superior fit to the data, as indicated by the reduced C-statistic values. Specifically, the 2T model yields a C-stat/dof of 4211.2/3691, while the 1T + IC model results in a C-stat/dof of 4230.1/3691. This difference in C-statistics suggests that the 2T model better captures the complexity of the spectrum across the entire energy range, offering a more accurate representation of the data than the 1T + IC model, even though the latter improves the fit relative to the single-temperature model alone. In other words, while the addition of the power-law component to the single-temperature model enhances the overall fit, our findings indicate that the two-temperature model is the statistically preferred model for describing the X-ray spectrum in its entirety.

To investigate whether the observed hard X-ray excess is linked to any of the point sources within the cluster, we identified point sources in the field of Abell 3667 using \textit{Chandra} data (Fig. \ref{fig: A3667_img}, right). By analysing the spectrum of the brightest X-ray source in the field, located at R.A. (J2000) = 303.14647 and Decl. (J2000) = -56.89514, we found that its spectrum is well described by a simple power-law model, yielding a photon index of \( 1.6 \pm 0.20 \). When we extrapolated the flux to the 20--80 keV energy range, we obtained a value of \( 2.22^{+1.54}_{-1.13} \times 10^{-13} \, \text{erg} \, \text{s}^{-1} \, \text{cm}^{-2} \), which is more than an order of magnitude lower than the measured hard X-ray excess. This significant difference effectively rules out the possibility that the excess is due to this point source.

Based on this, we conclude that the excess is most likely due to either a hot thermal component with a temperature of \( kT = 25.8^{+6.1}_{-2.1} \, \text{keV} \) or a non-thermal power-law component with a photon index of \( 1.6^{+0.1}_{-0.2} \). Among these models, the thermal component provides a statistically better fit to the X-ray spectrum, making it the preferred explanation for the observed hard X-ray excess in the cluster.

\section{Discussion}
\label{sec: disc} 
\subsection{Hard excess emission}
The detection of a hard X-ray excess in Abell 3667 has been the subject of extensive study, offering crucial insights into the physical processes shaping the ICM in dynamically active galaxy clusters. Early observations with \textit{BeppoSAX} \citep{Fusco-Femiano2001} and \textit{RXTE} \citep{Rephaeli2004} first revealed a significant X-ray excess above 10 keV, which was interpreted as non-thermal emission, typically modeled by a power-law spectrum. This excess was initially attributed to IC scattering of relativistic electrons by CMB photons. Relativistic electrons are expected to arise in dynamically active clusters undergoing mergers, where shocks and turbulence accelerate particles to relativistic speeds. Thus, the observed excess serves as key evidence for high-energy processes occurring in the ICM.

Subsequent \textit{Suzaku} observations \citep{nakazawa2009hard} confirmed the hard X-ray excess but introduced significant ambiguity in its interpretation. The data could be equally well described by two competing models: a thermal model with an exceptionally high temperature of $T = 19.2^{+4.7}_{-4.0}$ keV, or a non-thermal power-law component with a photon index $\Gamma = 1.4$. The thermal model suggests the existence of a highly energetic region within the ICM, possibly due to localized heating from merger shocks or adiabatic compression. In contrast, the non-thermal model implies the presence of relativistic electrons coexisting with the thermal plasma, potentially accelerated by shock waves or turbulent processes.

Further exploration by \citet{Ajello2010}, using combined \textit{XMM-Newton} and \textit{Swift}/BAT data, confirmed the hard X-ray excess in both Abell 3667 and the Bullet Cluster. For Abell 3667, they found that the data could be explained by either a hot thermal component with $T = 13.5^{+6.9}_{-2.2}$ keV or a non-thermal power-law component with $\Gamma = 1.8$, corresponding to a flux of $3.0^{+4.2}_{-0.7} \times 10^{-12} \,\mathrm{erg} \, \mathrm{s}^{-1} \, \mathrm{cm}^{-2}$ in the 50--100 keV range. These findings highlight the persistent degeneracy between thermal and non-thermal models, making it difficult to definitively disentangle the two components with current data.

If the hard X-ray excess is indeed non-thermal, it would provide compelling evidence for the presence of relativistic electrons and magnetic fields within the cluster. These electrons, coupled with magnetic fields, could produce synchrotron emission detectable at radio wavelengths. Indeed, Abell 3667 hosts prominent radio relics at its periphery \citep{Gasperin2022}, thought to trace merger-induced shocks capable of accelerating particles to relativistic energies. This aligns with the IC scattering interpretation of the hard X-ray excess, suggesting a close connection between the shock acceleration of particles and high-energy emission.

Alternatively, if the emission is thermal in origin, it would indicate the existence of an exceptionally hot region within the ICM. Such localised heating could result from merger-driven shocks or substructures within the cluster that have not yet reached thermal equilibrium. However, these extreme temperatures pose challenges for our understanding of the energy budget and heat distribution in the ICM, raising questions about the underlying microphysical processes responsible for energy transfer in the ICM.

\begin{figure}
\begin{center}
\includegraphics[width=1.0\columnwidth]{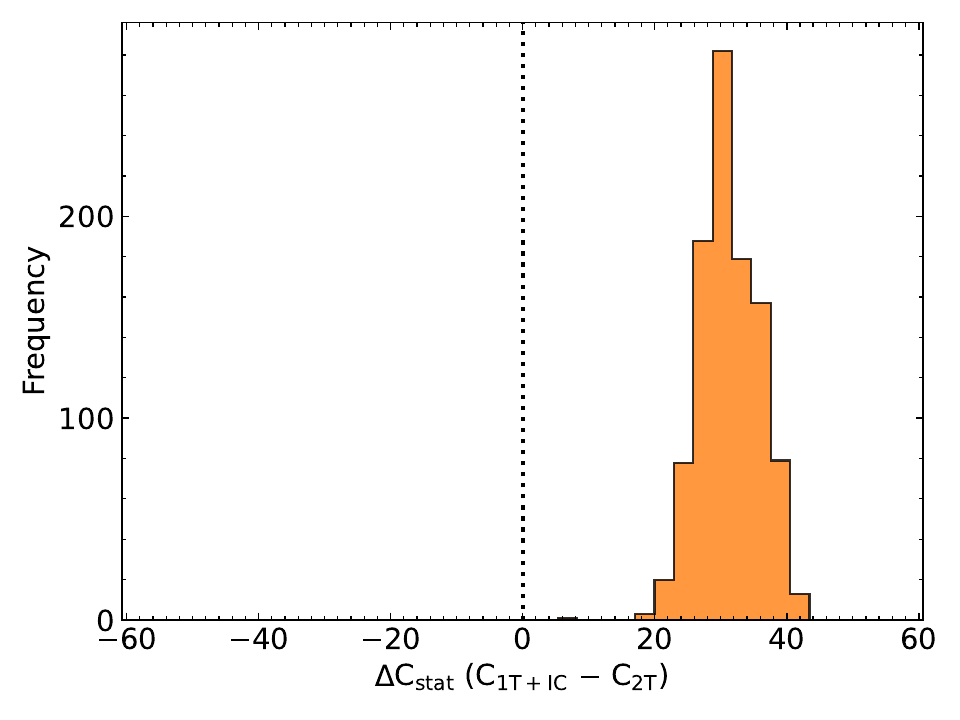}
\end{center}
\caption{Difference in C-stat values between the 1T + IC model (with a free photon index) and the 2T model, based on 1000 background iterations for each model. The dotted line marks the point where both models yield identical fit statistics to the spectral data. Values to the right of the line indicate a preference for the 2T model, while values to the left favor the 1T + IC model. In almost all cases, the 2T model is statistically preferred over the 1T + IC model with a free photon index.}
\label{fig: hist}
\end{figure}

In order to resolve this ambiguity, we analysed deep \textit{NuSTAR} observations of the central region of Abell 3667, which provide enhanced sensitivity at high energies and allow for more robust modelling of the hard X-ray spectrum. Our analysis shows that the hard X-ray excess is better described by a thermal model than by a non-thermal IC model. Specifically, the 2T thermal model yields a reduced C-statistic of 4211.2 for 3691 degrees of freedom, compared to the 1T + IC model, which gives a reduced C-statistic of 4230.1 for the same degrees of freedom. The lower C-statistic for the 2T model indicates a better overall fit, suggesting that the observed spectrum is dominated by thermal processes, rather than non-thermal IC scattering. Although the second component of the 2T model is very high (\(25.8_{-2.1}^{+6.1}\) keV), it is consistent with previous \textit{Suzaku} measurements \citep{nakazawa2009hard}. While a spatially resolved analysis could help clarify the origin of this high-temperature component, such an investigation is beyond the scope of the current work and will be addressed in future studies.

The preference for the 2T thermal model provides important constraints on the physical conditions within the cluster. The additional high-temperature component likely reflects a region of intense heating due to the merger shock, consistent with theoretical models of dynamically evolving clusters. However, we cannot fully rule out a non-thermal contribution, particularly given the presence of radio relics and the complex, turbulent environment of Abell 3667. Moreover, our \textit{NuSTAR} observations cover only the central region, and non-thermal emission may become more prominent in the outer regions, especially near the radio relics.

To assess the impact of systematic uncertainties in the background, we computed the difference in C-statistic values between the 2T and 1T + IC models using 1000 background realisations that incorporate systematic fluctuations, as described in \citet{Wik2014} and \citet{mirakhor2022possible}. The resulting distribution of C-stat differences, shown in Figure \ref{fig: hist}, reveals that in nearly all realisations, the 2T model is statistically preferred, suggesting it provides a better fit in most cases. Nevertheless, this does not entirely exclude a non-thermal interpretation. The data's limitation to the central cluster region means we may not fully capture contributions from the radio relics in the outskirts, which are thought to arise from non-thermal processes. These relics could significantly influence the overall emission profile and potentially alter the fit statistics. Therefore, while the 2T model is the best fit based on the available data, a more complex, non-thermal contribution may emerge when considering the outer regions of the cluster.


To fully resolve this ambiguity, deeper observations with \textit{NuSTAR} will be essential. Extended observations that cover regions outside the cluster core could provide valuable insights into the spatial distribution of the hard X-ray excess and help distinguish between thermal and non-thermal emission. In particular, the outskirts of Abell 3667, where prominent radio relics are observed, represent an ideal region for studying the non-thermal components of the ICM. By probing these outer regions, we could gain a better understanding of the contributions from both thermal X-ray emission and non-thermal processes, such as the synchrotron radiation from relativistic electrons and magnetic fields. High-resolution spectroscopy of these extended regions will be crucial for examining the nature of the hard X-ray emission and its connection to dynamic processes like merger shocks, turbulence, and the formation of radio relics, which are key features in clusters like Abell 3667. These deeper observations will offer new perspectives on how mergers shape and energise the ICM, providing more precise constraints on the physical processes that govern galaxy cluster evolution across cosmic time.

\subsection{Magnetic field strength}

To gain insights into the average magnetic field strength in Abell 3667, we can use the estimated flux of the non-thermal emission, under the assumption that this emission is produced by the same population of relativistic electrons responsible for the observed radio synchrotron emission. For a population of electrons with a power-law energy distribution, the magnetic field strength can be derived from the ratio of the radio synchrotron flux to the X-ray flux, as expressed in the following equation \citep{Govoni2004}:

\begin{equation}
\begin{aligned}
   B[\mu G]^{1+\alpha} &= h(\alpha) \frac{S_{\rm{R}}[{\rm{Jy}}]}{S_{\rm{X}}[{\rm{erg}}\, {\rm{s}}^{-1} \,{\rm{cm}}^{-2}]} (1+z)^{3+\alpha} \times \\
   &\quad (0.0545\nu_{\rm{R}}[{\rm{MHz}}])^\alpha \times \\
   &\quad \big(E_2[{\rm{keV}}]^{1-\alpha} - E_1[{\rm{keV}}]^{1-\alpha}\big),
\end{aligned}
\label{eq: magnetic_field}
\end{equation}

where \( h(\alpha) \) is a proportionality constant that depends on the electron spectral index \( \alpha \), \( S_{\rm{R}} \) is the radio flux density at the frequency \( \nu_{\rm{R}} \), and \( S_{\rm{X}} \) is the X-ray flux measured over the energy range \( E_1 \) to \( E_2 \). The spectral index \( \alpha \) reflects the energy distribution of the relativistic electrons.

Using the spectral index and the 20--80 keV flux density derived from the best-fit 1T + IC model (Section \ref{sec: spectra}), along with the radio flux density of \( 44 \pm 6 \, \mathrm{mJy} \) at 2.3 GHz \citep{Carretti2013}, we derive a lower limit on the magnetic field strength of \(  \sim 0.2 \, \mu \mathrm{G} \) for the central region of Abell 3667. This value is in good agreement with those found in other dynamically active galaxy clusters using the same approach, such as the Bullet Cluster \citep{Wik2014} and Abell 523 \citep{Cova2019}. Also, it is compatible with those derived from equipartition arguments and Faraday rotation measurements, which typically indicate magnetic field strengths on the order of a few \( \mu \mathrm{G} \) \citep[e.g.,][]{Kim1990, Bonafede2010}.

Note that the \( 44 \pm 6 \, \mathrm{mJy} \) measurement corresponds to the flux density of the entire radio bridge connecting the two relics, which extends beyond the NuSTAR field of view. However, since our NuSTAR observations cover the region of highest surface brightness within the bridge, this value likely captures the majority of the emission within the NuSTAR footprint. A more accurate magnetic field estimate for the specific region observed by NuSTAR would require a lower, spatially matched radio flux density, which would in turn reduce the inferred field strength. This emphasises the importance of spatially resolved, multi-wavelength observations to more precisely isolate and characterise the relevant emission components.



While the derived magnetic field strengths are consistent with previous estimates using the same method and compatible with equipartition-based estimates, future high-resolution observations will be crucial for more accurately constraining the magnetic field structure in Abell 3667 and other dynamically active clusters. Observations extending beyond the core, including the outskirts where prominent radio relics are found, will be particularly important for exploring the relationship between non-thermal emission, magnetic fields, and the state of the ICM. These deeper, multi-wavelength observations will enable more precise tests of the mechanisms that govern the amplification of magnetic fields in clusters and help refine our understanding of their role in the evolution of the ICM.

\section{SUMMARY}
\label{sec: summary}

The detection of a hard X-ray excess in Abell 3667 has been the subject of extensive investigation, revealing important insights into the physical processes occurring in the ICM of dynamically active galaxy clusters. Early observations suggested that the excess could be explained by non-thermal emission, likely from the IC scattering of relativistic electrons. However, subsequent observations with \textit{Suzaku} and \textit{XMM-Newton} introduced ambiguity, with the data being consistent with both thermal and non-thermal models.

Our analysis of \textit{NuSTAR} data, which covers the central region of Abell 3667, indicates that the hard X-ray excess in Abell 3667 is better described by a thermal model rather than a non-thermal IC model. The 2T thermal model, which includes an additional high-temperature component, provides a statistically better fit than the 1T + IC model. This suggests that the observed emission is primarily thermal in nature, likely due to localized heating from merger shocks. However, the possibility of a non-thermal contribution cannot be entirely excluded, particularly given the presence of radio relics in the cluster, which are thought to trace non-thermal processes.

Further investigations into the magnetic field strength in the central region of Abell 3667, using the ratio of radio synchrotron and X-ray fluxes, yield a lower limit on the magnetic field strength of \( \sim 0.2 \, \mu \mathrm{G} \). This result aligns with values found in other dynamically active clusters, and those derived from equipartition arguments and Faraday rotation measurements, which typically place the magnetic field strength at a few \( \mu \mathrm{G} \).


In summary, while the 2T thermal model provides a robust fit for the hard X-ray excess in Abell 3667, a more comprehensive analysis of the outer regions of the cluster is necessary to fully disentangle the thermal and non-thermal components. Observing the cluster outskirts presents significant challenges, which complicate the separation of emission from the thermal and non-thermal components. However, high-resolution, deep observations are crucial for overcoming these obstacles. Such observations will be pivotal for refining our understanding of the physical processes governing the ICM, particularly the roles of relativistic particles, magnetic fields, and turbulence in shaping cluster environments. Ultimately, these efforts will enhance our understanding of how mergers influence the evolution of galaxy clusters.

\paragraph{Acknowledgments}
We thanks the anonymous referee for the very helpful report. We acknowledge support from the NASA \textit{NuSTAR} grant 80NSSC23K1610. This research made use of data from the \textit{NuSTAR} mission, a project led by the California Institute of Technology, managed by the Jet Propulsion Laboratory, and funded by NASA. We thank the \textit{NuSTAR} Operations, Software, and Calibration teams for support with the execution and analysis of these observations. This research has made use of the NuSTAR Data Analysis Software \textsc{Nustardas} jointly developed by the ASI Science Data Center (ASDC, Italy) and the California Institute of Technology (USA).






\printendnotes

\defbibnote{preamble}{By default, this template uses \texttt{biblatex} and adopts the Chicago referencing style. However, the journal you’re submitting to may require a different reference style; specify the journal you're using with the class' \texttt{journal} option --- see lines 1--8 of \emph{sample.tex} for a list of options and instructions for selecting the journal.}

\printbibliography[prenote={preamble}]



\end{document}